\documentclass[12pt]{article}
\newcommand\Vus{{V_{us}}}
\newcommand\fplus{{f_+^{K\pi}(0)}}
\newcommand{\bi}{\begin{itemize}}
\newcommand{\ei}{\end{itemize}}
\usepackage{epsfig}
\usepackage{color}
\usepackage{rotating}
\usepackage{amssymb}

\def\beq{\begin{equation}}
\def\eeq#1{\label{#1}\end{equation}}
\def\eeqn{\end{equation}}

\def\beqa{\begin{eqnarray}}
\def\eeqa#1{\label{#1}\end{eqnarray}}
\def\eeqan{\end{eqnarray}}

\let\bar=\overbar

\def\Dslash{\not{\hbox{\kern-4pt $D$}}}
\def\dslash{\not{\hbox{\kern-2pt $\del$}}}

\def\msb{{\bar{\ssstyle M \kern -1pt S}}}

\def\Title#1{\begin{center} {\Large {\bf #1} } \end{center}}

\begin{document}

\Title{Kaon semi-leptonic form factor in lattice QCD\footnote{Proceedings of CKM 2012, the 7th International Workshop 
  on the CKM Unitarity Triangle, University of Cincinnati, 
  USA, 28 September - 2 October 2012}}

\bigskip\bigskip

%+\addtocontents{toc}{{\it D. Reggiano}}
%+\label{ReggianoStart}

\begin{raggedright}

{\it Andreas J\"uttner\index{J\"uttner, A.}\\
School of Physics and Astronomy\\
University of Southampton\\
Southampton, SO17 1AJ, UK\\
juettner@soton.ac.uk\\
  }
\abstract{This talk reviews recent lattice QCD simulations of the $K\to\pi$ 
semi-leptonic form factor.}\bigskip\bigskip
\end{raggedright}

\section{Introduction}
The unitary 
Cabibbo-Kobayashi-Maskawa matrix~\cite{Cabibbo:1963yz,Kobayashi:1973fv} 
(CKM) parameterises flavour changing processes in the Standard Model (SM).
We have reasons to believe that the SM will have to be extended by new 
physics contributions. This should also affect the flavour sector which 
we expect to manifest itself in terms of inconsistencies when 
overconstraining the parameters of the SM CKM matrix.
This talk reports on the status and ongoing improvements of 
determinations of the $K\to\pi$ semi-leptonic form factors which
are a crucial ingredient in the accurate and reliable determination of the 
CKM matrix element $|V_{us}|$. Due to the non-perturbative nature of
the form factor the only systematically improvable way to compute it 
is a simulation in lattice QCD.

%%%%%%%%%%%%%%%%%%%%%%%%%%%%%%%%%%%%%%%%%%%%%%%%%%%%%%%%%%%%%%%%%%%%%%%%%%%%%%%%
\section{Lattice determinations of $\fplus$}
The determination of $|\Vus|$ proceeds as follows: On the one hand,
one experimentally
measures the rate of  a flavour changing process 
$s\to u$, where $s$ is the strange quark
and $u$ the up quark. 
On the other hand one computes the SM
prediction for the same process which comprises contributions from 
electromagnetic, weak and strong interactions. 
Since the ultimate goal is a test of the SM any model-dependence should be
avoided and 
this is where progress in lattice simulations is currently being made.
The inclusion of electromagnetic and strong isospin breaking corrections in a 
fully non-perturbative fashion in lattice QCD is still in its infancy 
(see Tantalo's talk at this conference) but these effects will likely
soon be taken into account at the next level of 
sophistication of computations.
All results to date have been computed in pure
isospin symmetric ($m_u=m_d$)
lattice QCD~\cite{JLQCD11,RBCUKQCD10,RBCUKQCD07,Lubicz:2010bv,Lubicz:2009ht,Brommel:2007wn,Dawson:2006qc,Tsutsui:2005cj}
and the electromagnetic and isospin corrections are taken into account
only at the level of the analysis of the experimental 
data using chiral effective theory~\cite{Antonelli:2010yf} leading to 
% the following nocite includes the references to the work plotted in the
% scatter plot on f+
\nocite{Kastner:2008ch,Cirigliano:2005xn,Jamin:2004re,Bijnens:2003uy,Leutwyler:1984je} 
\begin{equation}\label{eqn:masterKtopi}
 |\Vus| \fplus=0.2163(5) \,.
\end{equation}
The remaining bit in the determination of $|V_{us}|$ is the prediction 
of $\fplus$ and 
table~\ref{tab:collaborations} summarises current activities which are all
based on simulations with either $N_f=2,\, 2+1$, or $2+1+1$ flavours of
dynamical quarks. With respect to the last edition of this workshop the 
number of collaborations working on $\fplus$ has increased. This is a positive
development in particular since the various efforts differ in their
approach and therefore systematic effects will likely 
contribute differently to individual results.

JLQCD, RBC/UKQCD and ETM calculate the form factor in terms of the 
QCD matrix element
\begin{equation}\label{eq:VME}
\langle \pi(p_\pi)|V_\mu|K(p_K)\rangle =
        { f_+^{K\pi}(q^2)}(p_K+p_\pi)_\mu
        + f_-^{K\pi}(q^2)(p_K-p_\pi)_\mu\,,
\end{equation}
of the vector current where $q^2=(p_K-p_\pi)^2$ is the momentum transfer. 
MILC instead  compute the form factor from the
scalar current matrix element,
\begin{equation}\label{eq:SME}
\langle \pi(p_\pi)|S|K(p_K)\rangle|_{q^2=0}=
	f_0^{K\pi}(0)\frac{m_K^2-m_\pi^2}{m_s-m_q}\,.
\end{equation}
Note that $\fplus=f_0^{K\pi}(0)$. The form factor as extracted from the
vector matrix element fulfils $\fplus=1$ in the SU(3)-symmetric limit 
also at finite lattice spacing which has the benefit of a symmetry-based 
suppression
of cut-off effects. In eq.~(\ref{eq:SME}) this symmetry is not manifest at 
finite lattice spacing and one expects
larger cut-off effects. In any case cut-off effects can be parameterised 
allowing to extrapolate observables to the continuum limit, provided that
results at different lattice spacings have been generated. 

In view of the 
determination of $|V_{us}|$ lattice QCD has to provide the form factors at 
the kinematical point $q^2=0$. 
In the finite lattice box hadron momenta assume discrete values
corresponding to Fourier modes as determined by the choice of boundary 
conditions. The kinematical point $q^2$ is therefore
naively (i.e. with periodic boundary conditions) not accessible. 
Partially twisted boundary conditions however allow to deal with this 
problem~\cite{Bedaque:2004ax,Sachrajda:2004mi,Boyle:2007wg,RBCUKQCD10} and 
by now all groups have adopted this technique.

Most current lattice simulations (except for MILC who have first simulation 
results with physical valence quark masses for $\fplus$)  
still simulate unphysically heavy $u$- and $d$-quarks
and results for $\fplus$ have to be extrapolated to the
physical point guided by chiral perturbation theory 
where the expansion of the form factor is around the SU(3)-symmetric limit, 
$\fplus=1+f_2+f_4+\ldots$ (also a small mistuning of the $s$-quark mass
can in this way be corrected).
In this expression $f_2$ is the NLO-term with the decay constant
as the only unknown~\cite{Gasser:1984ux,RBCUKQCD10} and 
$f_4$ is the NNLO-term~\cite{Bijnens:2003uy}. 

\begin{table}
{\footnotesize\noindent
\begin{tabular*}{\textwidth}[l]{@{\extracolsep{\fill}}lllllllll}
Collaboration & 
\hspace{0.15cm}\begin{rotate}{60}{published}\end{rotate}\hspace{-0.15cm}& $N_f$ &

\hspace{0.15cm}\begin{rotate}{60}{\parbox{1.8cm}{fermion formulation}}\end{rotate}\hspace{-0.15cm}&
\hspace{0.15cm}\begin{rotate}{60}{\parbox{1.8cm}{lightest pion mass}}\end{rotate}\hspace{-0.15cm}&
\hspace{0.15cm}\begin{rotate}{60}{technique}\end{rotate}\hspace{-0.15cm}
&
\hspace{0.15cm}\begin{rotate}{60}{$q^2$-dependence}\end{rotate}\hspace{-0.15cm}&
\hspace{0.15cm}\begin{rotate}{60}{$a$-dependence}\end{rotate}\hspace{-0.15cm}\\
&&&& \\[-0.3cm]
\hline
\hline&&&& \\[-3mm]
MILC      &&2+1+1&stag.& 135 &S&twbc&\checkmark\\[-3mm]
&&&& \\[-0.1cm]
\hline
&&&& \\[-0.3cm]
MILC      &&2+1  &stag.& 270 &S&twbc &\checkmark\\[-0mm]
JLQCD     &&2+1  &ovl. & 290 &V&twbc+interpol.&\\[-0mm]
RBC/UKQCD &\checkmark&2+1  &DWF  & 170 &V&twbc &\checkmark\\[-1mm]
&&&& \\[-0.3cm]
\hline
&&&& \\[-0.3cm]
ETM       &\checkmark&2    &TM   & 260 &V&twbc+interpol&\checkmark\\[-3mm]
 &&&& \\[-0.1cm]
\hline
\hline
\end{tabular*}}
\footnotesize
\begin{center}
\begin{minipage}[t]{.3\linewidth}
fermion formulations: \\[-7mm]
\bi
\item {{ stag.}} = staggered\\[-7mm]
\item {{ ovl.}} = overlap\\[-7mm]
\item {{ DWF}} = Domain Wall Fermions\\[-7mm]
\item {{ TM}} = twisted mass
\ei
\end{minipage}
\begin{minipage}[t]{.3\linewidth}
technique:\\[-7mm]
\bi
\item {{ V}} = direct computation of vector current\\[-7mm]
\item {{ S}} = computation via Ward identity
\ei
\end{minipage}
\begin{minipage}[t]{.3\linewidth}
$q^2$-dependence:\\[-7mm]
\bi
\item {{ twbc}} = twisted boundary conditions\\[-7mm]
\item {{ interpol.}} = interpolation in 
	momentum transfer to $q^2=0$
\ei
\end{minipage}
\end{center}
\caption{Summary of current efforts for computing $\fplus$ in lattice QCD.
}\label{tab:collaborations}
\end{table}

\begin{figure}
\footnotesize
\begin{center}
\hspace*{1mm}
\begin{minipage}[t]{.3\linewidth}
\begin{tabular}{lcl}
% \multicolumn{2}{l}{\mycite BMW Phys.Rev.D81:054507,2010}\\
 \hline\hline\\[-4mm]
 source &\multicolumn{2}{c}{ {$\delta\fplus$}}\\[-0mm]
 \hline\\[-4mm]
 statistical    &0.3\%\\[-1mm]
 chiral extrapolation   &\,0.4\%\\[-1mm]
 cont. extrapolation &0.1\%\\[-0mm]
 \hline\\[-4mm]
 total                  &0.5\%\\[-0mm]
 \hline\hline
 \end{tabular}
\end{minipage}\hspace*{1.8cm}
\begin{minipage}[c]{.60\linewidth}
\includegraphics[height=7cm]{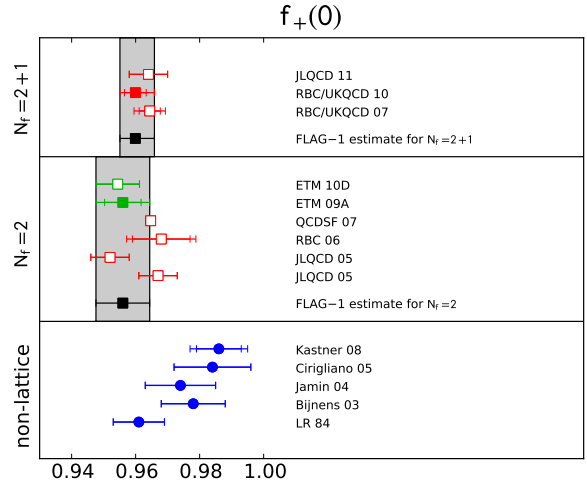}
\end{minipage}
\end{center}
\caption{Error budgets for state-of-the-art lattice computations for
$\fplus$ (left, RBC+UKQCD \cite{RBCUKQCD07,RBCUKQCD10})
and 
summary of current lattice and phenomenological 
results with the colour coding as in the FLAG report~\cite{Colangelo:2010et}.}\label{fig:results}
\end{figure}

The l.h.s. panel in figure~\ref{fig:results} shows a typical error budget
for $\fplus$ as determined from eq.~(\ref{eq:VME})
which underlines the importance of simulations at the physical point - 
the chiral extrapolation is by far the dominant source of systematic 
uncertainty.
\section{Conclusions}
The two largest systematic uncertainties in lattice computations of the 
$K\to\pi$ semi-leptonic form factor $\fplus$ are due to the interpolation 
of lattice data in the momentum transfer to $q^2=0$ and due to the extrapolation of lattice data in the quark mass. The former 
has now been removed in all simulations 
through the use of partially twisted boundary conditions which allow
for simulations directly at $q^2=0$. 
The latter is about to be removed by
simulating very close to or at the physical point. 
At the level of precision now reached in these computations
it will become important to incorporate also electro-magnetic and isospin 
effects into the simulation (Nazario Tantalo's talk at this conference)
eventually replacing the dependence on effective theory calculations.

\bigskip

{\bf Acknowledgements:}
The research leading to these results has received funding from the Euro\-pe\-an
Research Council under the European Union's Seventh Framework Programme
(FP7/2007-2013) / ERC Grant agreement 279757.

\bibliographystyle{JHEP}
\bibliography{CKM2012}
\end{document}